\documentclass[sigconf]{acmart}
\usepackage{etex}

\usepackage{subcaption} 

\acmConference[]{}

\usepackage{booktabs}   
\usepackage{subcaption} 
\usepackage[noend]{algpseudocode}

\usepackage{graphicx}
\usepackage{amsmath}
\usepackage{xspace}
\usepackage{footnote}
\usepackage{amsfonts}
\usepackage{natbib}
\usepackage{hhline}
\usepackage{diagbox}
\usepackage{multirow}
\usepackage{setspace}
\usepackage{epsfig}
\usepackage{url}

\usepackage{breakurl}
\usepackage{booktabs}

\usepackage{xcolor}
\usepackage{lipsum}
\usepackage{courier}
\usepackage{listings}
\usepackage{caption}
\usepackage{colortbl}
\usepackage{arydshln} 
\usepackage{makecell} 
\usepackage{tikz}
\usepackage{color}

\usepackage{rotating}

\usepackage{varwidth}
\usetikzlibrary{fit,arrows,calc,positioning,quotes}

\usepackage{pgfplots}

\usepackage{mathtools}
\usepackage{fancyvrb}
\usepackage{textcomp}

\usepackage{stmaryrd}

\usepackage{algorithm}
\usepackage[noend]{algpseudocode}

\usepackage{alltt}

\tikzset{
  -|-/.style={
    to path={
      (\tikztostart) -| ($(\tikztostart)!#1!(\tikztotarget)$) |- (\tikztotarget)
      \tikztonodes
    }
  },
  -|-/.default=0.5,
  |-|/.style={
    to path={
      (\tikztostart) |- ($(\tikztostart)!#1!(\tikztotarget)$) -| (\tikztotarget)
      \tikztonodes
    }
  },
  |-|/.default=0.5
}

\usetikzlibrary{shapes.geometric}

\long\def\com#1{}

\newcommand{\app}{{\sc Gmerge}\xspace}

\newcommand{\accuracy}{\ensuremath{64.6\%}\xspace}

\newcommand{\para}[1]{\smallskip\noindent {\bf #1}}

\newcommand{\squishlist}{
   \begin{list}{$\bullet$}
    { \setlength{\itemsep}{0pt}      \setlength{\parsep}{3pt}
      \setlength{\topsep}{3pt}       \setlength{\partopsep}{0pt}
      \setlength{\leftmargin}{3.5mm} \setlength{\labelwidth}{1em}
      \setlength{\labelsep}{0.5em} }
}

\newcommand{\squishend}{
    \end{list}  }

\def\BibTeX{{\rm B\kern-.05em{\sc i\kern-.025em b}\kern-.08em
    T\kern-.1667em\lower.7ex\hbox{E}\kern-.125emX}}

\begin{document}

\title{Can Pre-trained Language Models be Used to Resolve Textual and Semantic Merge Conflicts?}

\author{Jialu Zhang}
\affiliation{%
  \institution{Yale University}
  \city{New Haven}
  \state{Connecticut}
  \postcode{06511}
  \country{USA}
  }

 \author{Todd Mytkowicz}
\affiliation{%
  \institution{Microsoft Research}
  \city{Redmond}
  \state{Washington}
  \postcode{98052}
  \country{USA}
  }
  
  \author{Mike Kaufman}
\affiliation{%
  \institution{Microsoft Corporation}
  \city{Redmond}
  \state{Washington}
  \postcode{98052}
  \country{USA}
  }
  
 \author{Ruzica Piskac}
\affiliation{%
  \institution{Yale University}
  \city{New Haven}
  \state{Connecticut}
  \postcode{06511}
  \country{USA}
  }

 \author{Shuvendu K. Lahiri}
\affiliation{%
  \institution{Microsoft Research}
  \city{Redmond}
  \state{Washington}
  \postcode{98052}
  \country{USA}
  }

\begin{abstract}

	Program merging is standard practice when developers integrate their individual changes to a common code base.
When the merge algorithm fails, this is called a merge conflict. 
The conflict either manifests in \emph{textual merge conflicts} where the merge fails to produce code, or \emph{semantic merge conflicts} where the merged code results in compiler or test breaks.
Resolving these conflicts for large code projects is expensive because it requires developers to manually identify the sources of conflict and correct them.

In this paper, we explore the feasibility of automatically repairing merge conflicts (both textual and semantic) using {\it k-shot learning} with large neural language models (LM) such as GPT-3.
One of the challenges in leveraging such language models is to fit the examples and the queries within a small prompt (2048 tokens). 
We evaluate LMs and k-shot learning for two broad applications: (a) textual and semantic merge conflicts for a divergent fork Microsoft Edge, and (b) textual merge conflicts for a large number of JavaScript projects in GitHub. 
Our results are mixed: one one-hand, LMs provide the state-of-the-art (SOTA) performance on semantic merge conflict resolution for Edge compared to earlier symbolic approaches; on the other hand, LMs do not {\it yet} obviate the benefits of fine-tuning neural models (when sufficient data is available) or the design of special purpose domain-specific languages (DSL) for restricted patterns for program synthesis. 



\end{abstract}

\maketitle

\section{Introduction}
\label{sec:intro}

Merge conflicts today have been the common causes of broken pull requests, failure of continuous integration builds, and latent software defects in large projects~\cite{Ghiotto18}.
One of the key reasons for the trends in issues caused by merge conflicts is the increasing collaborative environment in large modern software.
A recent study showed that with thousands of people and tens of active branches committed to the same code base,
nearly 20 percent of all the merge attempts in the large project finally ended with a bad merge~\cite{Ghiotto18}.

A bad merge originates from either a \emph{textual} merge conflict or a \emph{semantic} merge conflict.  A textual merge conflict occurs when two developers differently edit the same line of code. Normally, a developer then must resolve the conflict manually (i.e., when using git merge).  In contrast, a semantic merge conflict occurs when there is no textual merge conflict but the merge results in a broken build, failing test (that would have otherwise passed), or unintended runtime behavior.  Such semantic merge conflicts cannot be automatically resolved and thus require manual fixes from developers~\cite{SungLKCW20}.  Such fixes greatly delay the development process.

Semantic merge conflicts can manifest in all forms of merge attempts, but we observed that 
they often appear in divergent forks. A divergent fork is created as a copy of the source
repository without the intention to 
rarely (if ever) merge back. The standard terminology refers to the source repository as the 
upstream project and a fork is called the downstream project. Downstream has an 
independent development history which is rarely merged back to the upstream~\cite{SungLKCW20}.
For example, Microsoft Edge, Opera and Brave are all based on the same upstream (Chromium).
Each downstream branch periodically pulls the latest updates from upstream and merges them 
with other branches in the downstream repository.
Using divergent forks saves a large amount of developer time by reusing 
functions and classes already defined and tested upstream, expedites the whole development process, and improves the maintainability of the code repository~\cite{SungLKCW20}.

\begin{figure*}[t!]
\centering\includegraphics[width=.75\linewidth]{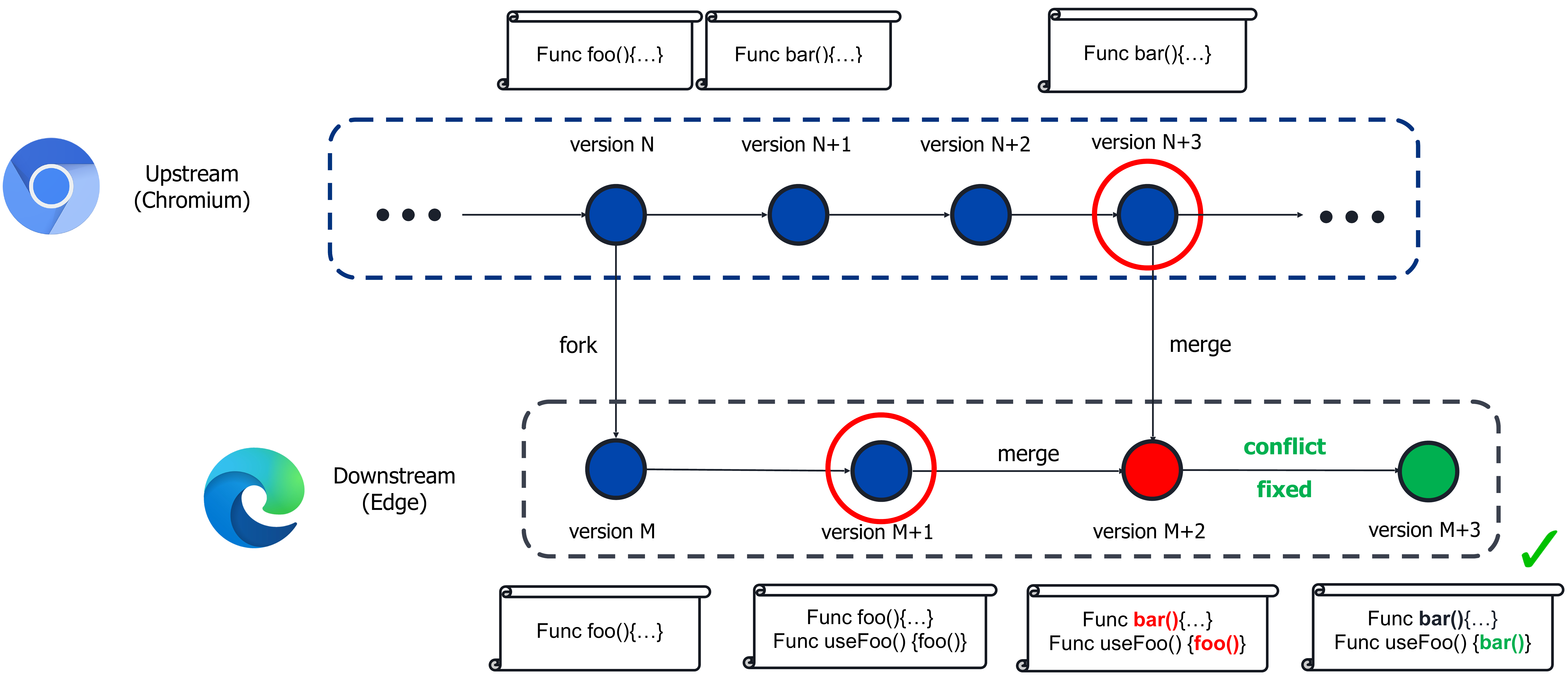}
\caption{An Example of Semantic Merge Conflict} 
\label{fig:foo}
\end{figure*}

However, in a divergent fork, downstream developers frequently suffer from both textual and semantic merge
conflicts. The reason is because upstream and downstream have an independent 
development processes with the downstream being out of sync with the upstream. Downstream developers often have little knowledge of current upstream updates.
Taking Microsoft Edge as an example, a study showed that over a three-month period there 
were more than 25,000 upstream commits in Chromium and they all had to be
merged downstream~\cite{SungLKCW20}. 
This level of merge frequency makes it difficult or even infeasible for downstream developers to inspect every upstream change before a merge. 

Figure~\ref{fig:foo} illustrates an example of a semantic merge conflict. A function \lstinline{foo()} was initially introduced upstream. 
After the original fork, another function \lstinline{useFoo()},
which invokes \lstinline{foo()}, was created downstream. 
Concurrently, in the upstream branch the function \lstinline{foo()} was renamed to \lstinline{bar()}. Later, during the next pull from upstream, there was no textual merge conflict between two versions circled in red. The definition of \lstinline{foo()} downstream did not change before 
the merge, so the merging algorithm simply renamed it into \lstinline{bar()}.
Thus, a semantic merge conflict is born, as \lstinline{foo()} now has no definition in the downstream and will not compile.

The example in Fig.~\ref{fig:foo} is not artificially contrived:  
semantic merge conflicts happen daily in the Microsoft Edge 
development process. A snippet illustrating the root cause of one such 
conflict is given in
Fig.~\ref{fig:edgeupstream}. It is a fragment of the log of
\lstinline{git diff} applied to two commits in the upstream branch. 
The snippet shows that function \lstinline{IsIncognito()} was renamed to \lstinline{GetIncognito()}.
After the merge, the downstream branch fails to compile with the  message, 
``no member named \lstinline{IsIncognito()} in \lstinline{BrowserView}''. 

\lstset{
numbers=left, 
numberstyle=\small, 
numbersep=8pt, 
frame = single, 
language=C}

{
\begin{figure}[!h]
    
\begin{lstlisting}[escapechar=!]
...
!\colorbox{pink!50}{- bool \textbf{IsIncognito()} const;}!
!\colorbox{green!10}{+ bool \textbf{GetIncognito()} const;}!
\end{lstlisting}
    \caption{Root cause of a semantic merge conflict in Chromium upstream}
    \label{fig:edgeupstream}
\end{figure}
}

To help a programmer deal with build breaks caused by semantic merge conflicts, we developed a tool, 
named \app, that automatically suggests merge conflict resolutions. \app takes as input 
a merge conflict and merge histories from both upstream and downstream.
\app returns a conflict resolution which indicates which lines of code need to
change, and how. 
In our 
particular Edge example, \app returns the conflict resolution that is given in Fig.~\ref{fig:edgedownstream}. It states that function 
\lstinline{IsIncognito()} should be renamed to function \lstinline{GetIncognito()}.

{
\begin{figure}[!h]
    
\begin{lstlisting}[escapechar=!]
if (!!browser &&
!\colorbox{pink!50}{-      ...GetBrowserViewForBrowser(browser)->\textbf{IsIncognito()})}!
!\colorbox{green!10}{+      ...GetBrowserViewForBrowser(browser)->\textbf{GetIncognito()})}!
    {return;}
    
\end{lstlisting}
    \caption{The resolution for the semantic merge conflict shown in Fig.~\ref{fig:edgeupstream} in Edge downstream.}
    \label{fig:edgedownstream}
\end{figure}
}

The motivation for this work 
stems from the need to have a tool that 
helps programmers in repairing semantic merge conflicts in Microsoft Edge. Repairing semantic merge conflicts 
is prohibitively expensive 
for downstream developers.
To find a root cause of the conflict, a developer needs to manually inspect the upstream 
commit history, which, for a large project, can be measured in thousands of commits. The 
previous study~\cite{SungLKCW20} have shown that through the course of three months 
around 800 commits were identified as attempts to solve merge conflicts. 
Each commit requires Edge developers additional time to resolve.



Our tool is based on {\it k-shot learning} with a large language model (GPT-3~\cite{gpt3}).
GPT-3 is a large language model that has been successfully deployed in many applications 
such as questions answering~\cite{gpt3}, text completion, source code generation~\cite{chen2021evaluating} and in many other fields.
The biggest difference between GPT-3 and other supervised machine learning 
models is that the user does not need to train the model specifically for their
application. The user only provides 
a few "shots" (or examples to prime the model) as input to GPT-3 and GPT-3 achieves competitive results,  
compared to other supervised machine learning 
models. A {\emph{shot}} is a standard term that describes a question/answer pair.
Motivated by GPT-3's successful applications in other fields, this paper investigates using 
GPT-3 to resolve merge conflicts. A k-shot learning approach with GPT-3 has significant engineering benefits as it does not require expensive task specific (in our case merge is the task) fine-tuning.  Further, k-shot learning implies we can use existing large language models and leverage them as they get better.

There are two main challenges with a k-shot approach: data curation and prompt engineering.
Data curation automatically extracts source code changes related to
merge conflicts. These changes are extracted from both upstream and downstream commits
and they are converted into an intermediate representation (IR). 
Prompt engineering takes data in the IR format and 
translates the data into input consumable by GPT-3. 
A key challenge with prompt engineering is that GPT-3's input is limited to 2048 tokens and thus the shot and query must fit within it.  To tackle this challenge, \app applies string pattern analysis and heuristics in prompt engineering.

Finally, to empirically evaluate our approach, we have run \app on real-world Microsoft Edge semantic merge conflicts.
Our evaluation shows (we use the developer's actual fixes as the ground truth), \app learns the correct resolutions at the state-of-the-art (SOTA) \accuracy of accuracy.
Our evaluation demonstrates the effectiveness of k-shot learning, which provides a cost-effective and language-agnostic solution for real-world semantic merge conflicts.

We then generalize our approach to textual merge conflicts and evaluate the 
effectiveness of our data curation and prompt engineering in this domain on two case studies (the first based on program synthesis and the second based on fine-tuning large scale language models).  Both of these existing tools require significant engineering while a k-shot approach is relatively simpler.
In the first, we show k-shot learning has performance on par with current SOTA tools, while
in the second, SOTA tools significantly outperform \app. However, note that 
in both case studies, the current SOTA approaches require special purpose 
domain-specific languages (DSL) for program synthesis or fine-tuned machine 
learning models.
When we compare the performance \app and those tools without fine-tuning, \app 
outperforms them.

In summary, we make the following contributions:
\begin{itemize}

\item We present a data-driven tool \app that uses k-shot learning with a large language model (GPT-3) to automatically find repairs for merge conflicts.
\item We propose a method of prompt engineering that translates merge conflict examples and queries to a small prompt for GPT-3.
\item We perform an evaluation of \app on both textual and semantic real-world merge conflicts problems. We obtained the state-of-the-art (SOTA) performance on semantic merge conflict resolution for a divergent fork, and comparable performance on textual merge 
conflicts problems including divergent forks and a large number of JavaScript projects in GitHub.
\end{itemize}

\if

However, the downstream branch and the upstream branch merged at version \lstinline{M + 2} in the downstream, and function definition was changed from \lstinline{foo()} to \lstinline{bar()}. However, the call site of the function \lstinline{foo()} in the downstream commit remains the same before and after merge, making Clang fire an error that \lstinline{foo()} has no definition. To fix this build break, developers need to look back to the upstream commit history to understand where the definition of the function \lstinline{bar()} comes from and change all the call site of function \lstinline{foo()} to \lstinline{bar()} as the version \lstinline{M + 3} shows. 
Fixing such a build break is non-trivial because: (1) The root cause of the build break is introduced in upstream commits instead of downstream. Since two branches develop individually, the developers in the Edge development team have no prior knowledge of the changes in the upstream commits.
(2) Updates in the upstream branch happen frequently, developers often need to look for thousands of commits in the upstream to locate where the error is, wasting much developers' time.

In modern software development, fork and merge back is the common practice.
It allows developers to make changes to the code at the same time for different purposes and later integrate all the changes for a merged stable version.
However, merge conflicts happen when merging algorithm such as Git fails to automatically resolve differences among merged sources.
It contains two cases, textual conflicts and non-textual conflicts. 
First, Git fails to generate a file with no textual conflict. 
This is mainly caused by two merged sources change the same line to different contents.
It often needs human labor to resolve this kind of case.
Second, Git is able to generate such files with no textual conflict.
However, the merged code cannot be correctly compiled; Running the merged code will produce a compiler error.
The second case of semantics merge conflicts is also called build breaks.

Build breaks happen frequently in large software development, divergent fork specifically. In a divergent fork, a fork (downstream) is created at some starting point from the source repository (upstream) to have an independent development and barely merged back to the upstream~\cite{SungLKCW20}.
For example, Microsoft Edge, Opera and Brave are all based on the same upstream (Chromium).
They each periodically pull the latest updates from the Chromium, merge with the downstream repository and customize functions based on their own needs.
This design saves massive developers' time by reusing functions and classes that were already tested by the upstream, expedites the whole development process and improves the maintainability of the whole code repository.

However, non-textual merge conflicts happen frequently when the current downstream commit merged with the latest upstream.
For example, in Fig.~\ref{fig:foo}, at version \lstinline{M} in the downstream commit, a function definition \lstinline{foo()} was initially forked from the upstream commit. Later in the downstream, another function \lstinline{useFoo()} was created in version \lstinline{M + 1} to customize the needs in Edge development. At the same time, the upstream branch the function \lstinline{foo()} was renamed to \lstinline{bar()}. Since two branches have individual developments, so before the merge happen, there is no error in both of code.
However, the downstream branch and the upstream branch merged at version \lstinline{M + 2} in the downstream, and function definition was changed from \lstinline{foo()} to \lstinline{bar()}. However, the call site of the function \lstinline{foo()} in the downstream commit remains the same before and after merge, making Clang fire an error that \lstinline{foo()} has no definition. To fix this build break, developers need to look back to the upstream commit history to understand where the definition of the function \lstinline{bar()} comes from and change all the call site of function \lstinline{foo()} to \lstinline{bar()} as the version \lstinline{M + 3} shows. 
Fixing such a build break is non-trivial because: (1) The root cause of the build break is introduced in upstream commits instead of downstream. Since two branches develop individually, the developers in the Edge development team have no prior knowledge of the changes in the upstream commits.
(2) Updates in the upstream branch happen frequently, developers often need to look for thousands of commits in the upstream to locate where the error is, wasting much developers' time.

\fi

\section{System Overview}
\label{sec:overview}

\begin{figure}[t]
  \centering
   \resizebox{1\linewidth}{!}{
  \begin{tikzpicture}[thick,scale=0.8, every node/.style={scale=0.65}]

\tikzstyle{a} = [ellipse,  draw, color=black, fill=gray!30, inner sep=0.2cm, align=center]
\tikzstyle{b} = [rectangle, draw, node distance=0.6cm, minimum width=1.5cm, minimum height=2em, thick, align=center, inner sep=0.2cm]
\tikzstyle{c} = [fill=gray!30, rounded corners=5pt, inner sep=0.2cm, align=center]
\tikzstyle{decision} = [diamond, draw, aspect=2, align=center, inner sep=3pt, spacing, fill=blue!5!white]
    
    \node [b] (Conflict) {Failed \\ Merge};
    \node [b, left=.5cm of Conflict] (Up) {Upstream \\ Changes};
    \node [b, right=.5cm of Conflict] (Down) {Downstream \\ Changes};
    \node [c, below=.5cm of Conflict] (Data) {Data Curation};
    \node [b, below=.5cm of Data] (IR) {Conflict Description};
    
    \node [c, right=0.9cm of IR] (PromptMethod) {Prompt Engineering};
    \node [b, right=0.9cm of PromptMethod] (PromptInput) {Prompt};
    \node [a, above=.5cm of PromptInput] (GPT3) {GPT-3};
    \node [b, above=.5cm of GPT3] (Resolution) {Conflict Resolution};
    
    \draw [->] (Up.south) -- (Data.west);
    \draw [->] (Down.south) -- (Data.east);
    \draw [->] (Conflict.south) -- (Data.north);
    \draw [->] (Data.south) -- (IR.north);
    \draw [->] (IR.east) -- (PromptMethod.west);
    \draw [->] (PromptMethod.east) -- (PromptInput.west);
    \draw [->] (PromptInput.north) -- (GPT3.south);
    \draw [->] (GPT3.north) -- (Resolution.south);

\end{tikzpicture}
  }
  \caption{\app Overview}
  \label{fig:overview}
\end{figure}

The overview of the \app tool is shown in Fig.~\ref{fig:overview}.  It
takes as input three parameters: a merge conflict, and commits in the
upstream and downstream repositories that constitute the merge. As
output, \app returns 
a merge conflict resolution.

\app contains two main modules: data curation and prompt engineering.

The data curation module takes as input the upstream and downstream
commits along with the downstream semantic merge conflict including
the compiler error messages. It generates JSON files as the
intermediate representation (IR) for the prompt engineering module.
We call this file a {\emph{conflict description}}.  This paper
contains three case studies, but data curation is extensively used
only in the first case study.  However, when we run \app on an
existing benchmark where the conflicts have already been curated, \app
skips this module and directly goes to the next module. 

The purpose of the prompt engineering module is to translate a
conflict description file into the small prompt format required as
input for GPT-3.  The prompt engineering is a technical module and for
each of our case studies we needed to apply various different
heuristics, described in more details in
Sections~\ref{subsec:semanticprompt} and~\ref{sec:textualmerge}.  This
is due to the fact that conflict description files for each case study
have different format.

In the next two sections, we describe how we apply \app{} to different
merge conflict case studies.

\section{Case Study 1: Resolving Semantic Merge Conflicts for A Divergent Fork}
\label{sec:semanticmerge}

\subsection{Data Collection and Curation}
\label{subsec:analyzer}
Merge conflicts in downstream forks are often introduced by commits
made in upstream repository, so the goal of the data collection and
curation process is to identify and extract all the source code
changes in upstream that are related to the given merge conflict.  In
real-world software development, this is also the first step that the
programmer manually performs in trying to resolve a failed merge.  For
large upstream repositories such as Chromium (of which Microsoft Edge
is a downstream divergent fork), searching through the thousands of
upstream commits is a tedious and error-prone task.  In \app we
automate this whole process.

\app takes the commit logs of the repository, the semantic merge
conflicts including the compiler error messages as the inputs, and
outputs a JSON file for each semantic merge conflict.  This JSON file
contains all code-level changes relevant to the target merge conflict.
The files are designed to be self-contained, in the sense that it is
sufficient to check the JSON file to gain all the information relevant
for the given semantic merge conflicts.  Each file has three key
components: 1) the set of relevant changes from upstream, 2) the
conflict to be fixed in downstream, and 3) the resolution provided by
the real-world developers.  The third component is only relevant for
establishing the ground-truth for our evaluation.  To evaluate how
accurate are conflict resolutions suggested by \app, we evaluated it
on existing semantic merge conflicts. We store the manual repair so
that we can compare it to our derived solution.

Fig.~\ref{fig:jsonexample} shows an example JSON file that \app
generated for the merge conflict given in Fig.~\ref{fig:edgeupstream}
and Fig.~\ref{fig:edgedownstream}.
In Fig.~\ref{fig:jsonexample}, line 2 to line 12 are the relevant changes from upstream.
Line 13 is the merge conflict and line 14 is the resolution provided by the Edge developers.

{
\begin{figure}[!h]
\begin{lstlisting}[]
{
    "UpstreamChanges": [
        {
            "Before": "const bool incognito = browser_view_->IsIncognito();",
            "After": "const bool incognito = browser_view_->GetIncognito();"
        },
        {
            "Before": "bool IsIncognito() const;",
            "After": "bool GetIncognito() const;"
        }
        ...
    ],
    "DownstreamConflict": "BrowserView::GetBrowserViewForBrowser(browser)->IsIncognito())",
    "DownstreamFix": "BrowserView::GetBrowserViewForBrowser(browser)->GetIncognito())"
}
\end{lstlisting}
    \caption{The JSON example generated by \app for merge conflicts shown in Fig.~\ref{fig:edgeupstream} and Fig.~\ref{fig:edgedownstream}.}
    \label{fig:jsonexample}
\end{figure}
}

Semantic merge conflicts happen when the code produced by a merge
(either through the textual merge or through a user resolution) cannot
be compiled; the compiler fails to produce the executable code and
outputs an error message. We use this error message to identify the
relevant keywords and prune the space of source code changes in
upstream based on these keywords.  Taking
Fig.~\ref{fig:edgedownstream} as an example, Clang results in the
error message ``Cannot find the definition for function
\lstinline{IsIncognito()}.''
This is a typical error message for a semantic merge conflict due to function renaming.
From this message, \app identifies \lstinline{IsIncognito()} as the keyword relevant to this error message.

\app automatically extracts the keywords for each semantic merge
conflict; it starts by extracting the strings in the quotation marks
in the error message.  To be conservative, if the keyword has a C++
class specifier in it as the prefix, e.g.,
\lstinline{browser::IsIncognito()}, we collect both its type specifier
\lstinline{browser} and the function name \lstinline{IsIncognito()} as
the keywords.  These keywords are used as the seed for further search
in upstream commits.  \app extracts every deleted line in the upstream
code diff that contains the keywords, and it additionally extracts the
line following immediately after the deleted line.  This line
typically contains the new name of the renamed symbol.  As this change
may span multiple lines, in \app we use a heuristic that collects all
the lines until we reach an unchanged or deleted line.

For each semantic merge conflict manifested in downstream, we extract
the line of code that has a compiler error.  To obtain the
ground-truth user resolution for such a merge conflict, we identify
the repair commit in downstream that follows the conflict that 1)
changed the same line in the same file as the semantic merge conflict
and 2) the compiler error for that specific merge conflict no longer
existed after that commit.  Similar to upstream, we collect the repair
code region in downstream fork in the same way.


\subsection{Prompt Engineering}
\label{subsec:semanticprompt}

The prompt engineering module applies various heuristics on the JSON
representation of the conflicts to translate merge conflict examples
and queries to succinct prompts for GPT-3.  The output of GPT-3 is the
resolution of the merge conflict.

\para{Prompt Structure} Each prompt is a question to GPT-3 and it has
two parts: shots and one query.  The first part, the shots is optional
in the prompt.  If the prompt contains no shot, it is called zero-shot
learning.

Each shot itself is an example of how a previous merge conflict is
resolved.  We use shots to provide a context to GPT-3 about the
current task, and provide an example of what should be the correct
form of the output.  Fig.~\ref{fig:simpleone-shotlearning} shows an
example of one shot learning. By providing an example to GPT-3, the
model learns the relation between apple to red, and then applies the
relation to the eggplant, and outputs purple as the result.

{
\begin{figure}[!h]
    
\begin{lstlisting}[escapechar=!]
Question: Apple?
Answer: Red

Question: Eggplant?
Answer:
\end{lstlisting}
    \caption{An Example of one shot learning in GPT-3. The output of this example by GPT-3 is Purple. Line 1 and line 2 are the shot. Line 4 is the query.}
    \label{fig:simpleone-shotlearning}
\end{figure}
}

In our application, we use shots to let the GPT-3 model better
understand how the merge conflict is solved and what the ideal
resolution looks like.  Fig.~\ref{fig:one-shotlearning} is the
real-world example on how we use \app to resolve the merge conflict in
Fig.~\ref{fig:edgedownstream}. 

{
\begin{figure}[!h]
    
\begin{lstlisting}[]
Question:
--web_app_info->app_url = url;
++web_app_info->start_url = url;

-web_app_info->app_url = url;

Answer:
+web_app_info->start_url = url;

Question:
--if (browser_view_->IsIncognito()||!browser_view_->IsBrowserTypeNormal())
++if (browser_view_->GetIncognito()||!browser_view_->GetIsNormalType())

--bool IsIncognito() const;
++bool GetIncognito() const;

- ...GetBrowserViewForBrowser(browser)->IsIncognito()) 

Answer:
+ ...GetBrowserViewForBrowser(browser)->GetIncognito()) 
\end{lstlisting}
    \caption{An Example of merge conflict resolution using one shot learning in GPT-3. The shot is represented in line 1 to line 8. The query starts from line 10 and it ends at line 17. The line 20 is not in the prompt, and it is the output of GPT-3 model for the resolution of the merge conflict.}
    \label{fig:one-shotlearning}
\end{figure}
}

\para{Prompt Format} In the prompt, the lines start with a double \lstinline{--} and \lstinline{++} represent the conflict related changes in upstream.
The line starting with a single \lstinline{-} appears only at the end of the query and it represents the merge conflict in downstream.
The line starting with a single \lstinline{+} is not in the prompt. Instead, it is the output of GPT-3 for the resolution of the merge conflict.


\para{Prompt Content} GPT-3 has a fixed input size, 2048 tokens.  This
is in sharp contrast to the massive code diffs of thousands of
commits.  A single JSON file could contain thousands of lines of code
changes related to a merge conflict.  To leverage the power of GPT-3,
one of the key challenges in \app is to fit the examples and the
queries into this small prompt.

We adopt a heuristic to ensure that we have prioritize diverse
representation of ``UpsteamChanges'' (in the JSON file) in the prompt.
Our intuition is that we want each pair of ``UpsteamChanges'' has
distinct string \lstinline{edit sequence}.  
Each \lstinline{edit sequence} is a list of operations that are applied to strings.
Applying the \lstinline{edit sequence} on the first string produces
the second one.  There are three operations in the edit, the addition
\lstinline{+}, the deletion \lstinline{-} or the equivalence
\lstinline{=}.  For any string character replacement, we used
\lstinline{-+} to represent it.  We also omit the space padding in our
\lstinline{edit sequence} definition.

We use the Python library \lstinline{difflib.ndiff} to compute the
string difference.  For simplicity, if any two adjacent characters
have the same operation, we will only keep one operation in the final
\lstinline{edit sequence}.  For example, given line 14 and line 15 as
the input, the \lstinline{edit sequence} is \lstinline{=-+=}.  \app
makes sure that every \lstinline{--} and \lstinline{++} string pairs
in the prompt has a distinct \lstinline{edit sequence} pattern.  In
this way, we managed to fit all the shots and the queries into the
small prompt (2048 tokens).  In Sec.~\ref{subsec:semanticeval}, we
demonstrate the effectiveness of our prompt content design.

\subsection{Evaluation}
\label{subsec:semanticeval}

We evaluate the efficacy of \app on semantic merge conflicts by answering the following questions:
\begin{enumerate}
\item Can \app resolve semantic merge conflicts in divergent forks?
\item Does prompt engineering positively affects the accuracy?
\item Are larger language models more accurate than smaller ones?
\end{enumerate}


\para{Experiment Setup} 
We collected Edge merge conflicts from Aug 2020 to April 2021.  We
used four clang compiler error types to collect those conflicts that
are related to the semantic merge conflict. These four error types are
\lstinline{err_no_member}, \lstinline{err_no_member_suggest},
\lstinline{err_undeclared_var_use_suggest} and
\lstinline{err_undeclared_use_suggest}.
We obtained a total of 379 semantic merge conflicts for our empirical
evaluation.  Each conflict is prepossessed into a JSON file format.

The evaluation metric in \app is that if the actual fix by users is a
prefix of our generated resolution, we consider this resolution is
correct.  We take this evaluation metric because GPT-3 is an
auto-regressive model, which sometimes outputs the string without a
stop unless the output reaches a fixed length.  In addition, an actual
user fix could cross multiple lines.  In \app, we extracted only the
first line as the user fix.  Therefore, we required the actual fix by
users should be the prefix of our resolution.

This evaluation metric is considered a conservative one because there
might exist multiple correct ways for users to fix a merge conflict.
All experiments were conducted on a Windows computer with an Intel i7
CPU and 48 GB of RAM.




\para{Can \app resolve semantic merge conflicts in divergent forks?}
Table~\ref{table:res_baseline} shows the performance of \app on our
dataset. \app has an overall accuracy of \accuracy after ten model
trials.  Table~\ref{table:res_baseline} also includes a comparison of
\app to three baselines.
We first compared \app to a heuristic-based approach,
\lstinline{StringMerge}.  And then, we compared \app to the
state-of-the-art string-based program synthesis
approach~\cite{verbruggen2021semantic},
\lstinline{Transformation.Text}.  And then we evaluated how the
choices of language model (GPT-3 and GPT-J) affected the results.

Our first baseline \lstinline{StringMerge}, is a heuristic-based
approach designed by empirically analyzing patterns in semantic merge
conflicts for a divergent fork.  For each before-after code change
pair, \lstinline{StringMerge} computes the patterns of string diff
character by character.  Then it applies the learned pattern in
downstream to generate a fix.  It mimics the way Edge developers
manually fix such a merge conflict in the real world.
Table~\ref{table:res_baseline} shows \lstinline{StringMerge}’s
performance. \app performs better in terms of resolution accuracy
(\accuracy vs 30.1\%).

Our second baseline is the state-of-the-art program synthesis approach
on string transformation \cite{verbruggen2021semantic}.  \app performs
much better in terms of resolution accuracy (\accuracy vs 25.9\%) and
surprisingly even the heuristic-based \lstinline{StringMerge}
outperforms it.  One possible reason is that
\lstinline{Transformation.Text} is a generic string transformation
tool, so the pattern in semantic merge conflict resolution is too
complex for \lstinline{Transformation.Text} to learn.
Fig.~\ref{fig:motiv_edgeupstream} is such a challenging example that
is difficult for the existing program synthesis approach to resolve.

Our third baseline \app (GPT-J) is introduced to evaluate how the size
of the language model affects the result.  GPT-3 and GPT-J have
similar architectures, but GPT-3 has 175 billion parameters while
GPT-J has 6 billion.  Our evaluation shows that the size of the model
affects the ability to resolve semantic merge conflicts.  \app
performs much better than \app (GPT-J) in terms of resolution accuracy
(\accuracy vs 39.1\%) \app (GPT-J) has performed better than our
heuristic based approach \lstinline{StringMerge} and
\lstinline{Transformation.Text}.  This shows that resolving semantic
merge conflicts is a non-trivial problem, and a large language model
is able to automatically generate a resolution for semantic merge
conflicts with high accuracy.

\begin{table}[]
\caption{Evaluation of \app with baselines on merge conflict resolution accuracy.}
\label{table:res_baseline}
\resizebox{\linewidth}{!}{
\begin{footnotesize}
\begin{tabular}{|c|c|c|c|c|}
\hline
{\color[HTML]{000000} }         & {\color[HTML]{000000} \app (GPT-3)}            & \app (GPT-J)           & StringMerge  & Transformation.Text    \\ \hline
{\color[HTML]{000000} Accuracy} & {\color[HTML]{000000} \textbf{64.6\% (245/379)}} & 39.1\% (148/374) & 30.1\% (114/379) & 25.9\% (98/379) \\ \hline
\end{tabular}
\end{footnotesize}
}
\end{table}

\para{Does prompt engineering positively affects the accuracy?}
Table~\ref{table:res_gpt3} shows how prompt engineering affects the
accuracy of merge conflict resolution in \app.  One of the advantages
of GPT-3 is that it only needs a few examples (shots) as the input.
We evaluated \app in two prompt structures: one shot and zero shot.

\begin{table}[t]
\caption{Evaluation for different prompt designs in \app on Resolving Edge Semantic Merge Conflicts.}
\label{table:res_gpt3}
\begin{footnotesize}
\begin{tabular}{|l|c|c|c|l}
\cline{1-4}
{\color[HTML]{000000} }          & \multicolumn{1}{l|}{{\color[HTML]{000000} First Pair }} & \multicolumn{1}{l|}{{\color[HTML]{000000} \vtop{\hbox{\strut Maximal Test}\hbox{\strut (without heuristics)}}}} & \multicolumn{1}{l|}{{\color[HTML]{000000} \bf{\vtop{\hbox{\strut Maximal Test}\hbox{\strut (with heuristics)}}} }} &  \\ \cline{1-4}
{\color[HTML]{000000} One shot}  & {\color[HTML]{000000} 44.1\%}                                           & {\color[HTML]{000000} 60.4\%}                                                 & {\color[HTML]{000000} \textbf{64.6\%}}                                        &  \\ \cline{1-4}
{\color[HTML]{000000} Zero shot} & {\color[HTML]{000000} 33.2\%}                                           & {\color[HTML]{000000} 35.1\%}                                                 & {\color[HTML]{000000} 40.0\%}                                                 &  \\ \cline{1-4}
\end{tabular}
\end{footnotesize}
\end{table}

One of the major technical challenges in \app is to fit the shot and
the query into the prompt in GPT-3, we have evaluated \app on three
different prompt structures in Table.~\ref{table:res_gpt3}.  ``First
pair'' means that we only choose the first conflict related source
code change in the JSON file to form the query.  ``Maximal Test
(without heuristics)'' takes as many changes as possible in the
original sequence until the size of the prompt becomes larger than
2048 tokens.  ``Maximal Test (with heuristics)'' takes the heuristics
described in Sec.~\ref{subsec:semanticprompt} as the filtering method
to prioritize some changes in the prompt.

The evaluation shows that having a shot as the input to the language
model significantly improves the results in all prompt structures.
This meets our expectation because having a shot not only clearly
pinpoints the current task to the model but also provides an example
of what is the expected output from the model.  Moreover, the
evaluation shows that providing more conflicts related code changes as
the context improves the accuracy of the model.  It further shows that
with the heuristics, \app achieved the highest accuracy of \accuracy.

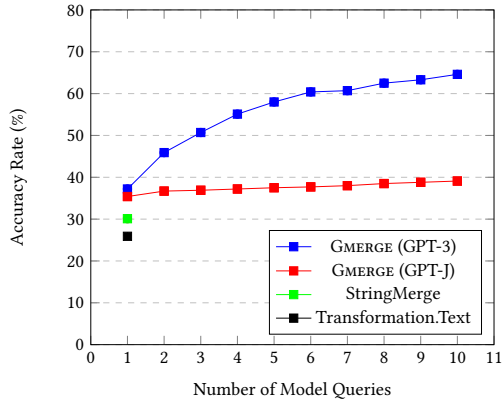
\begin{figure}[t]
 \centering
      \resizebox{.8\linewidth}{!}{\begin{tikzpicture}
\begin{axis}[
    title={},
    xlabel={Number of Model Queries},
    ylabel={Accuracy Rate (\%)},
    xmin=0, xmax=11,
    ymin=0, ymax=80,
    xtick={0,1,2,3,4,5,6,7,8,9,10,11},
    ytick={0,10,20,30,40,50,60,70,80},
    legend pos=south east,
    ymajorgrids=true,
    grid style=dashed,
]

\addplot+[only marks,
    error bars/.cd,
    y dir=both,
    y explicit,
    error bar style={line width=1pt},
    error mark options={
      rotate=90,
      blue,
      mark size=4pt,
      line width=0.5pt
    }
    ]
    coordinates {
    
    (1,37.2)
    (2,45.9)
    (3,50.7)
    (4,55.1)
    (5,58.0)
    (6,60.4)
    (7,60.7)
    (8,62.5)
    (9,63.3)
    (10,64.6)

    };
    
\addplot+[only marks,
    error bars/.cd,
    y dir=both,
    y explicit,
    error bar style={line width=1pt},
    error mark options={
      rotate=90,
      white,
      mark size=4pt,
      line width=0.5pt
    }
    ]
    coordinates {
    
    (1,35.4)
    (2,36.7)
    (3,36.9)
    (4,37.2)
    (5,37.5)
    (6,37.7)
    (7,38.0)
    (8,38.5)
    (9,38.8)
    (10,39.1)

    };

\addplot+[only marks,
    error bars/.cd,
    y dir=both,
    y explicit,
    error bar style={line width=1pt},
    error mark options={
      rotate=90,
      white,
      mark size=4pt,
      line width=0.5pt
    }
    ]
    coordinates {
    
    (1,30.1)

    };    

\addplot+[only marks,
    error bars/.cd,
    y dir=both,
    y explicit,
    error bar style={line width=1pt},
    error mark options={
      rotate=90,
      white,
      mark size=4pt,
      line width=0.5pt
    }
    ]
    coordinates {

    (1,25.9)

    };    

\addplot[
    color=blue,
    mark=square*,
    ]
    coordinates {
    
    (1,37.2)
    (2,45.9)
    (3,50.7)
    (4,55.1)
    (5,58.0)
    (6,60.4)
    (7,60.7)
    (8,62.5)
    (9,63.3)
    (10,64.6)
    };
    
\addplot[
    color=red,
    mark=square*,
    ]
    coordinates {
    
    (1,35.4)
    (2,36.7)
    (3,36.9)
    (4,37.2)
    (5,37.5)
    (6,37.7)
    (7,38.0)
    (8,38.5)
    (9,38.8)
    (10,39.1)
    };

\addplot[
    color=green,
    mark=square*,
    ]
    coordinates {
    
   (1,30.1)
   
    };
    
\addplot[
    color=black,
    mark=square*,
    ]
    coordinates {
    
   (1,25.9)

    };
    \legend{,,,,\app(GPT-3),\app(GPT-J),StringMerge,Transformation.Text}
    
\end{axis}
\end{tikzpicture}}
      \caption{Accuracy Comparison for \app on Resolving Semantic Merge Conflicts}
      \label{fig:accuracy_comparison}
\end{figure}

GPT-3 and GPT-J each output one resolution at one model trial.  In our
experiment, we repeatedly query GPT-3, and if the resolution is
produced in any of the trials, we mark the merge conflict as
``resolved''.  We evaluated how the number of trials affect the model
accuracy.  Fig.~\ref{fig:accuracy_comparison} shows that the overall
accuracy of GPT-3 and GPT-J both increased with the number of model
trials For example, for GPT-3, ten independent trials achieves the
accuracy of \accuracy in contrast to the accuracy of 37.2\% with only
one trial.  Compared to the GPT-3 model, we only observed a modest
accuracy gain in the GPT-J model.  \lstinline{StringMerge} and
\lstinline{Transformation.Text} have no accuracy gain because they
produce a deterministic result in every run.

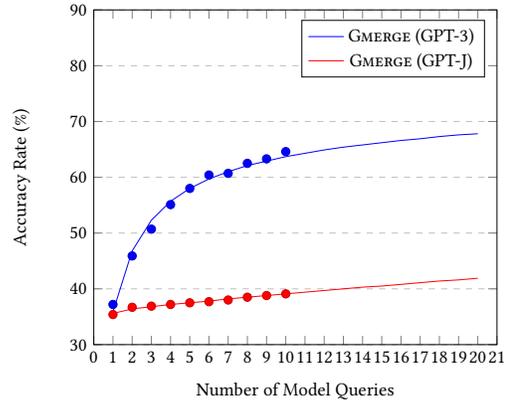
\begin{figure}[t]
 \centering
      \resizebox{.8\linewidth}{!}{\begin{tikzpicture}

\begin{axis}[
    title={},
    xlabel={Number of Model Queries},
    ylabel={Accuracy Rate (\%)},
    xmin=0, xmax=21,
    ymin=30, ymax=90,
    xtick={0,1,2,3,4,5,6,7,8,9,10,11,12,13,14,15,16,17,18,19,20,21},
    ytick={30,40,50,60,70,80,90},
    legend pos=north east,
    ymajorgrids=true,
    grid style=dashed,
]

    \addplot[
        scatter,%
        scatter/@pre marker code/.code={%
            \edef\temp{\noexpand\definecolor{mapped color}{rgb}{\pgfplotspointmeta}}%
            \temp
            \scope[draw=mapped color!80!black,fill=mapped color]%
        },%
        scatter/@post marker code/.code={%
            \endscope
        },%
        only marks,     
        mark=*,
        point meta={TeX code symbolic={%
            \edef\pgfplotspointmeta{\thisrow{RED},\thisrow{GREEN},\thisrow{BLUE}}%
        }},
    ] 
    table {
    x   y   RED   GREEN   BLUE
    1   37.2    0 0 1
    2   45.9    0 0 1
    3   50.7    0 0 1
    4   55.1    0 0 1
    5   58.0    0 0 1
    6   60.4    0 0 1
    7   60.7    0 0 1
    8   62.5    0 0 1
    9   63.3    0 0 1
    10  64.6    0 0 1
    1 35.4  1 0 0
    2 36.7 1 0 0
    3 36.9 1 0 0
    4 37.2 1 0 0
    5 37.5 1 0 0
    6 37.7 1 0 0
    7 38.0 1 0 0
    8 38.5 1 0 0
    9 38.8 1 0 0
    10 39.1 1 0 0
    }
    ;

    \addplot[
    color=blue,
    mark=none,
    ]
    coordinates {
    
(1,35.9)
(2,46.8)
(3,52.3)
(4,55.7)
(5,58.0)
(6,59.7)
(7,61.0)
(8,62.1)
(9,62.9)
(10,63.7)
(11,64.3)
(12,64.9)
(13,65.4)
(14,65.8)
(15,66.2)
(16,66.6)
(17,66.9)
(18,67.3)
(19,67.6)
(20,67.8)
    };

    \addplot[
    color=red,
    mark=none,
    ]
    coordinates {
    
(1,35.6)
(2,36.4)
(3,36.8)
(4,37.2)
(5,37.5)
(6,37.8)
(7,38.2)
(8,38.5)
(9,38.8)
(10,39.1)
(11,39.4)
(12,39.7)
(13,40.0)
(14,40.3)
(15,40.5)
(16,40.8)
(17,41.1)
(18,41.4)
(19,41.6)
(20,41.9)
    };
 \legend{,\app(GPT-3),\app(GPT-J)}
 \end{axis}
\end{tikzpicture}}
      \caption{Curve fitting for the simulated data vs. observed data. Two fitting curves are represented by line. The observed data are shown in sparse spots in the graph.}
      \label{fig:fit}
\end{figure}

\para{Are larger language models more accurate than smaller ones?}

One benefit of \app is that its k-shot approach does not require
expensive task specific fine-tuning.  Thus, \app can benefit from
large scale language autoregressive models.  In this section, we
demonstrate that the size of the model has a significant impact on
\app's task specific accuracy.  Fig.~\ref{fig:accuracy_comparison}
shows that the overall accuracy of GPT-3 increased more sharply than
GPT-J with the increase of model queries.  Approximately 30\% of the
additional merge conflicts are resolved if we query the GPT-3 model
multiple times.  In contrast, for GPT-J, only 5\% of the additional
merge conflicts can be resolved in this setting.


Based on what we observed from the result, we come up with the
following interesting hypothesis: Given a merge conflict, GPT-J is
likely to have a resolving probability close to zero (non-solvable) or
close to one (definitely solvable).  In other words, the density of
the merge conflicts, which can be resolved after several model trials,
is much higher in GPT-3 than GPT-J.

To validate this hypothesis, for both GPT-3 and GPT-J, our goal is to
use a high degree polynomial function to continuously model the
probability density of the samples in terms of different conflict
resolving probabilities.  We specifically focused on the proportion of
samples that have resolving probability close to zero (non-solvable)
or close to one (definitely solvable).  We used the gradient descent
algorithm~\cite{gradientdescent} to minimize the loss function in our
training.  The loss function was computed by the sum of the squared
errors in our setting.


With our learned probability density function, we simulated the
accuracy graph of the first twenty trials of the GPT-3 and the GPT-J
respectively to see how our simulated graph fits the observed results.
Fig.~\ref{fig:fit} shows that our function closely fits our observed
results.

\subsection{A Challenging Example Resolved by \app}
\label{subsec:motiv}

In this subsection, we illustrate the complexity of real-world merge
conflicts that downstream developers face in daily development.
We closely inspect the example of a broken build taken from the Edge
repository. For example, in Fig.~\ref{fig:motiv_edgedownstream}, the
compiler error message only indicated that it could not find a
definition for \lstinline{PermissionRequestType}.

The correct resolution to this particular problem is given the line
annotated with + in Fig.~\ref{fig:motiv_edgedownstream}. This was a
repair that the developer has committed.

{
\begin{figure}[!h]
    
\begin{lstlisting}[escapechar=!]
!\colorbox{pink!50}{- PermissionRequestType::PERMISSION\_CAMERA\_PAN\_TILT\_ZOOM:}!
!\colorbox{green!10}{+ RequestType::kCameraPanTiltZoom:}!
    
\end{lstlisting}
    \caption{A solution for the challenging semantic merge conflict resolution in Edge downstream.}
    \label{fig:motiv_edgedownstream}
\end{figure}
}

To correctly resolve such a merge conflict, developers need to learn
how the upstream context changed and then apply the similar changes to
the downstream context.  To derive this particular resolution, it was
not enough to find the relevant file in the commit history and then
propagate the changes: the developer needed to find three different
files and manually combine the changes described in those files to
derive the required resolution. We list the most relevant changes in
Fig.~\ref{fig:motiv_edgeupstream}.

{
\begin{figure}[!h]
    
\begin{lstlisting}[escapechar=!]
!\colorbox{pink!50}{- permissions::PermissionRequestType:: PERMISSION\_NOTIFICATIONS}!
!\colorbox{green!10}{+ permissions::RequestType::kNotifications}!

!\colorbox{pink!50}{- permissions::PermissionRequestType request\_type}!
!\colorbox{green!10}{+ permissions::RequestType request\_type}!

!\colorbox{pink!50}{- permissions::PermissionRequestType::PERMISSION\_NOTIFICATIONS)}!
!\colorbox{green!10}{+ if (request\_type == permissions::RequestType::kNotifications)}!



\end{lstlisting}
    \caption{Root cause of a challenging semantic merge conflict in Chromium upstream.}
    \label{fig:motiv_edgeupstream}
\end{figure}
}

The developers first detected that the root cause of this compiler
error was due to fact that \lstinline{PermissionRequestType} has been
renamed to \lstinline{RequestType} in upstream.  However, applying
these changes to the \lstinline{PermissionRequestType} downstream
still did not resolve this merge conflict.  This was due to the fact
that \lstinline{PERMISSION_NOTIFICATIONS} in upstream was changed to
\lstinline{kNotifications}.  After the developer learned how the
upstream context changed and applied similar changes to the
downstream context, the semantic merge conflict is resolved by
changing \lstinline{PERMISSION_CAMERA_PAN_TILT_ZOOM} to
\lstinline{kCameraPanTiltZoom}.

The existing merge conflict resolution
approaches~\cite{SungLKCW20,verbruggen2021semantic,PanLNGLK21} are not
helpful in such cases, because their learning algorithms are limited
when it comes to learning and combining complex string transformation.
Indeed, in our empirical evaluation, given in
Sec.~\ref{subsec:semanticeval}, we show that none of our baseline
methods could resolve this merge conflict.  With the prompt shown in
Fig.~\ref{fig:challengingexample}, \app generated line 25 as the
resolution to this conflict.

{
\begin{figure}[!h]
    
\begin{lstlisting}[]
Question:
--web_app_info->app_url = url;
++web_app_info->start_url = url;

-web_app_info->app_url = url;

Answer:
+web_app_info->start_url = url;

Question:
...

--"request", permissions::PermissionRequestType::PERMISSION_NOTIFICATIONS,
++"request", permissions::RequestType::kNotifications, requesting_origin);

--permissions::PermissionRequestType request_type,
++permissions::RequestType request_type,

--permissions::PermissionRequestType::PERMISSION_NOTIFICATIONS) {
++if (request_type == permissions::RequestType::kNotifications) {

- case permissions::PermissionRequestType::PERMISSION_CAMERA_PAN_TILT_ZOOM:

Answer:
+ case permissions::RequestType::kCameraPanTiltZoom:
\end{lstlisting}
    \caption{A challenging merge conflict example resolved by \app. Line 1 to line 8 are the shot, line 10 to line 22 are the query to GPT-3.}
    \label{fig:challengingexample}
\end{figure}
}

\subsection{Discussion}
\label{subsec:semanticdiscussion}

\para {Automation Level} The closest existing work to \app is the
MrgBldBrkFixer~\cite{SungLKCW20}, which investigated the feasibility
of automatically fixing the semantic merge conflicts in Microsoft
Edge.  It computes the differences of Abstract Syntax Trees (ASTs) of
the two programs to identify the changes in upstream for a given
symbol and then applies such a patch on the conflict in downstream for
a fix.

However, MrgBldBrkFixer is not fully automated, contrary to \app. It
needs downstream developers to manually classify the semantic
conflicts and assigns type for these conflicts. Therefore,
MrgBldBrkFixer still requires manual labor in resolving merge
conflicts. This is the reason why we cannot use MrgBldBrkFixer as the
baseline in our evaluation.

Automation is important in resolving semantic merge conflicts for a
divergent fork because the motivation of our work is that it often
costs developers much time to manually identify the source of
conflicts and correct them.

\para {Is GPT-3 able to generate out-of-vocabulary resolution for
  semantic merge conflicts?} If the resolution of the conflict
contains a token that is not contained in the prompt, it is an
out-of-vocabulary resolution.  It is difficult even for 
experienced developers to obtain such an out-of-vocabulary resolution.
We then automatically check how many times the actual resolution
contains tokens that are not in the prompt.

We found that 286 out of 379 cases, the resolution is in the
vocabulary of the given prompt.  GPT-3 gets the answer for 221 of them
with 77.3\% accuracy.

Surprisingly, for the rest of 93 examples, GPT-3 gets 14 correct
resolutions (15.1\%.)  Fig.~\ref{fig:challengingexample} is such an
example.  The details of our result is in Table~\ref{table:feasible},
which shows GPT-3 is able to generate complex merge conflict
resolutions.  \begin{table}[ht!]
\centering
\caption{\app generates resolution even if the resolution contains a token that is out of the input vocabulary.}
\label{table:feasible}
\resizebox{\linewidth}{!}{

\begin{footnotesize}
\begin{tabular}{|c|c|c|}
\hline
{\bf Tokens Out of Vocabulary} & {\bf \app resolved the conflict} & {\bf Occurrences}  \\ \hline
Not Required          & Yes     & 221     \\ \hline
Not Required           & No    & 65    \\ \hline
Required          & Yes     & 14     \\ \hline
Required          & No    & 79    \\ \hline
\end{tabular}
\end{footnotesize}
}
\end{table}


\subsection{Threats to Validity}
\label{subsec:validity}
\app cannot provide any guarantee to the resolution of merge
conflicts.  \app relied on Clang compiler message to locate the
conflict related changes in upstream.  Therefore, if the compiler
fails to pinpoint where the error is or the git commit history is not
accurate, \app cannot handle such merge conflicts.

\app cannot resolve such a merge conflict that is resolved by simply
deleting the line that contains merge conflict.  Based on our
expertise, this is a rare case in day to day development in a
divergent fork, and should not be recommended in the standard
practice.

\section{Generalization \app on Textual Merge Conflicts}
\label{sec:textualmerge}

To generalize \app to resolving textual merge conflicts, we conducted two case studies, 1) textual merge conflicts for the divergent fork Microsoft Edge, and 2) textual merge conflicts for a large number of JavaScript projects in GitHub. 
In both studies, the repositories that contain the merge conflicts are not available directly in the benchmark, thus making \app infeasible to do the data curation.
We run \app on the two existing benchmarks~\cite{dinella2021deepmerge,PanLNGLK21} to resolve merge conflicts using only the prompt engineering module.

\subsection{Case study 2: \app on resolving textual merge conflicts for a divergent fork}
\label{subsec:textualprose}

\para{Textual Merge Conflict} Fig.~\ref{fig:edge_textual_motiv} shows an example of a real-world textual merge conflict in Edge development.
The conflict is caused by the header file \lstinline{url_utils.h} in the forked branch that is an alternate version of the header file \lstinline{google_utils.h} in the main branch.
To resolve this issue, downstream developers kept the one in the forked branch and excluded the one in the main branch.
Running \app on this example will automatically produce the same resolution.

{
\begin{figure}[!h]
    
\begin{lstlisting}[escapechar=!]
Main branch(Chromium):
!\colorbox{pink!50}{- \#include ``components/google/core/common/google\_util.h'' }!
!\colorbox{green!10}{+ \#include ``components/variations/net/omnibox\_http\_headers.h'' }!
Forked Branch (Edge):
!\colorbox{green!10}{+ \#include ``components/microsoft/core/common/url\_utils.h'' }!
!\colorbox{green!10}{+ \#include ``components/variations/edge\_features.h'' }!
\end{lstlisting}
    \caption{A challenging example for resolving a textual merge conflict for a divergent fork. The line that starts with a ``+'' means this line is kept in the final merged code. The line that starts with a ``-'' means this line is removed from the final merged code.}
    \label{fig:edge_textual_motiv}
\end{figure}
}

\para{Benchmark Description} The problem of textual merge conflict resolution for header files and Macro in large projects has been studied in the paper~\cite{PanLNGLK21} and the benchmark is publicly available.
For each merge conflict in the benchmark, it shows how a textual merge conflict is resolved in Microsoft Edge development.
Each conflict is either a C++ header file conflict or a Marco related conflict, and the size of the merge conflict has up to two lines of changes.
This benchmark is collected from Microsoft Edge development repository in an eight-week period (March 30 2020 to April 24, 2020).
In total, this benchmark has 122 textual merge conflicts due to header file conflicts and 38 conflicts due to Macro conflicts.

\para{Prompt Engineering} We use the same prompt structure as in prior case studies.
However, without the assistance of data curation module, in this case study, we can only pick the examples in the existing benchmark to form the shots and use each of unused examples as query to evaluate \app.

We have two prompt engineering methods here.
First, we randomly select a few examples from each category to form the shot.
We ended up with five header file examples and two Marco examples in the shot.
We name this prompt engineering method as the ``Randomly selected shots''.
Second, we use our domain specific knowledge to pick two typical examples as the shot.
We name this prompt engineering method as the ``Representative shots''.

{
\begin{figure}[!h]
    
\begin{lstlisting}[escapechar=!]
Question:
<<<<<<< HEAD
#include "chrome/browser/ui/views/accessibility/caption_bubble_controller_views.h"
=======
#include "chrome/browser/ui/views/accessibility/hc_with_theme_bubble_view.h"
>>>>>>>

Answer:
#include "chrome/browser/ui/views/accessibility/caption_bubble_controller_views.h"
#include "chrome/browser/ui/views/accessibility/hc_with_theme_bubble_view.h"

Question:
<<<<<<< HEAD
#include "base/notreached.h"'
=======
#include "base/metrics/histogram_macros.h"'
>>>>>>>

Answer:
#include "base/metrics/histogram_macros.h"'
#include "base/notreached.h"'

Question:
<<<<<<< HEAD
#include "build/build_config.h"
=======
#include "media/media_buildflags.h"
>>>>>>>

Answer:
\end{lstlisting}
    \caption{An Example of Prompt in Textual Merge Conflict Resolution by \app}
    \label{fig:prose_gpt_example}
\end{figure}
}

For example, in Fig.~\ref{fig:prose_gpt_example}, line 1 to line 10 is our shot, which has two header file merge conflicts and their resolutions. 
Our target conflict is shown from line 23 to line 28. By feeding the shot to the GPT-3 model, it outputs:
\begin{lstlisting}
#include "build/build_config.h"
#include "media/media_buildflags.h"
\end{lstlisting}
and this is the exact resolution provided by the Edge developers.

\para{Evaluation and Discussion} Table~\ref{table:res_prosemerge} shows the accuracy of resolution on Edge header file and Macro textual merge conflict dataset.
Compared to the existing work, which requires a careful design of domain specific language, our tool \app has better accuracy on Macro related textual merge conflicts. 
For header file related merge conflicts, \app has a modest accuracy of 60.0\%.
This is mainly because \app does not have the domain specific knowledge for the repository in the input, which is used to resolve header file related merge conflicts.
Fig.~\ref{fig:edge_prose_logging1} is such an example that can only be resolved by using prior domain specific knowledge.

{
\begin{figure}[!h]

\begin{lstlisting}[escapechar=!]
Main branch(Chromium):
!\colorbox{green!10}{+ \#include ``base/notreached.h'' }!
Forked Branch (Edge):
!\colorbox{pink!50}{- \#include ``base/logging.h'' }!
!\colorbox{green!10}{+ \#include ``base/mojom/scoped\_native\_library.h'' }!
\end{lstlisting}

    \caption{A merge conflict that cannot be resolved without prior repository domain specific knowledge. The ``base/logging.h'' should always be removed from the forked branch because the forked branch, Microsoft Edge uses a different logging system. }
    \label{fig:edge_prose_logging1}
\end{figure}
}

To resolve such a merge conflict, the existing solution~\cite{PanLNGLK21} crafted a new domain-specific language that captures the patterns from historical data as resolution strategies, and used program synthesis to learn such repeated resolutions. 
Applying the learned strategies to the new unseen merge conflicts will automatically synthesize a resolution.
However, without access to the historical data of the full repository, the following `` always deleting \lstinline{logging.h} this pattern cannot be inferred by \app.

\begin{table}
\caption{Merge conflict resolving accuracy for \app on Edge header file and Macro textual merge conflict dataset.}
\label{table:res_prosemerge}
\resizebox{\linewidth}{!}{

\begin{footnotesize}
\begin{tabular}{|l|c|c|c|}
\hline
{\color[HTML]{000000} } & {\color[HTML]{000000} SOTA~\cite{PanLNGLK21}} & {\color[HTML]{000000} Randomly Selected Shots} & {\color[HTML]{000000} Representative Shots}          \\ \hline
{\color[HTML]{000000} Header File}   & {\color[HTML]{000000} 91.8\%(112/122)} & {\color[HTML]{000000} 49.6\% (58/117)}       & {\color[HTML]{000000} 60.0\% (72/120)} \\ \hline
{\color[HTML]{000000} Macro } & {\color[HTML]{000000} 94.4\%(35/38)}  & {\color[HTML]{000000} 100\% (36/36)}  & {\color[HTML]{000000} 100\% (36/36)}     \\ \hline
\end{tabular}

\end{footnotesize}
}
\end{table}

\subsection{Case study 3: \app on textual merge conflicts for a large number of JavaScript projects in GitHub.}
\label{subsec:textualdeepmerge}

\app is not just limited to divergent fork development infrastructure.
In this section, we evaluate \app to resolve textual merge conflicts collected over a large collection of open-source GitHub repositories. 

We evaluated \app on an existing Javascript merge conflicts benchmark used in DeepMerge~\cite{dinella2021deepmerge}.

\begin{table}
\caption{A merge conflict instance in the Deepmerge benchmark~\cite{dinella2021deepmerge}. Variant A and B are two versions of code changed from Base O. Resolution R is the fix provided by the developers. Merge size is computed by adding the lines of A and B.}
\label{table:deepmerge_example}
\begin{footnotesize}
\resizebox{1\columnwidth}{!}{
\begin{tabular}{@{}cccc@{}}
\toprule
{\color[HTML]{000000} \begin{tabular}[c]{@{}c@{}}Base O\\ (base.js)\end{tabular}}               & {\color[HTML]{000000} \begin{tabular}[c]{@{}c@{}}Variant A\\ (a.js)\end{tabular}}                                     & \begin{tabular}[c]{@{}c@{}}Variant B\\ (b.js)\end{tabular} & \begin{tabular}[c]{@{}c@{}}Resolution R\\ (r.js)\end{tabular}                     \\ \midrule
{\color[HTML]{000000} \begin{tabular}[c]{@{}c@{}}var b = 5.7;\\ var y = floor(b);\end{tabular}} & {\color[HTML]{000000} \begin{tabular}[c]{@{}c@{}}let b = x + 5.7;\\ var y = floor(b);\\ console.log(y);\end{tabular}} & var y = floor(x + 5.7);                                    & \begin{tabular}[c]{@{}c@{}}var y = floor(x + 5.7);\\ console.log(y);\end{tabular} \\ \bottomrule
\end{tabular}
}
\end{footnotesize}
\end{table}

\para{Benchmark Description} This benchmark~\cite{dinella2021deepmerge} contains thousands of real-world textual merge conflict examples collected from Github. 
Table~\ref{table:deepmerge_example} shows such an example.
Resolving merge conflicts in this benchmark is challenging because the conflicts  1) are collected from various open-source repositories other than a single repository as in the case of Edge, so the benchmark contains more generic code patterns and 2) vary in terms of their merge sizes. Some examples have large merge size up to hundreds of lines.

\para{Prompt Engineering} 
Similar to case study 2, without the assistance of the data curation module, we picked the examples in the existing benchmark to form the shots, and used the rest of the examples as queries to evaluate \app.

For this case study,  we chose the same shot content but with different shot formats.
We adopted two formats of shot designs shown in Fig.~\ref{fig:deepmerge_shot}.
Our goal is to check that if using different shot formats improves the final resolution accuracy.

{
\begin{figure}[!h]

        \begin{minipage}{.46\columnwidth}
	\begin{lstlisting} [frame = None,  caption={Classic Conflict Formulation},mathescape=true]
Questions:
<<<<<<<
let b = x + 5.7
var y = floor(b)
console.log(y)
|||||||
var b = 5.7;
var y = floor(b);
======= 
var y = floor(x + 5.7);
>>>>>>>

Answers:
var y = floor(x + 5.7);
console.log(y);
	\end{lstlisting}
    \end{minipage}
    \begin{minipage}{0.46\columnwidth}
	\begin{lstlisting} [frame = None, caption={Merge Tuple Formulation},mathescape=true]
Questions:
a.js:
let b = x + 5.7
var y = floor(b)
console.log(y)
base.js:
var b = 5.7;
var y = floor(b);
b.js:
var y = floor(x + 5.7);


Answers:
var y = floor(x + 5.7);
console.log(y);

        \end{lstlisting}
    \end{minipage}
    
    \caption{Two formats of shot designs.}
    \label{fig:deepmerge_shot}
\end{figure}
}

\para{Evaluation and Discussion} Compared to the state-of-the-art~\cite{dinella2021deepmerge}, \app has a suboptimal performance when the input merge size is small (less than eight lines). However, when the input size is large (greater than eight lines), \app outperformed the SOTA.

\begin{table}
\caption{Merge Conflict Resolving Accuracy for \app on Deepmerge~\cite{dinella2021deepmerge} Dataset. The first row is the input size of the merge. The prompt format that \app uses is on a par with the naive input representation for Deepmerge.}
\label{table:res_deepmerge}
\begin{footnotesize}
\resizebox{1.\columnwidth}{!}{
\begin{tabular}{|l|c|c|c|c|c|c|}
\hline
{\color[HTML]{000000}   }                                      & {\color[HTML]{000000} {[}0,3{]} lines}    & {\color[HTML]{000000} {[}4,5{]} lines}       & {\color[HTML]{000000} {[}6,7{]} lines}      & {\color[HTML]{000000} {[}8,10{]} lines}   & {\color[HTML]{000000} {[}\textgreater{}10{]} lines}  & {\color[HTML]{000000} Overall} \\ \hline
{\color[HTML]{000000} \app (tuple style)} & {\color[HTML]{000000} 29.1\% }  & {\color[HTML]{000000} 14.9\% }     & {\color[HTML]{000000} 12.46\% }   & {\color[HTML]{000000} 10.2\%}     & {\color[HTML]{000000} 2.48\% }    & {\color[HTML]{000000} 11.99\%}          \\ \hline
{\color[HTML]{000000} \app (conflict style)} & {\color[HTML]{000000} 32.11\% } & {\color[HTML]{000000} 17.52\% } & {\color[HTML]{000000} 12.86\% } & {\color[HTML]{000000} 11.96\% } & {\color[HTML]{000000} 4.46\%}    & {\color[HTML]{000000} 15.02\%}     \\ \hline
{\color[HTML]{000000} Deepmerge(Top 3,naive)} & {\color[HTML]{000000} N/A} & {\color[HTML]{000000} N/A} & {\color[HTML]{000000} N/A} & {\color[HTML]{000000} N/A} & {\color[HTML]{000000} N/A}     & {\color[HTML]{000000} 14.09\%}     \\ \hline
\multicolumn{1}{|c|}{{\color[HTML]{000000} Deepmerge(Top 1,tuned)}}          & {\color[HTML]{000000} 78.40\%}            & {\color[HTML]{000000} 56.50\%}               & {\color[HTML]{000000} 37.04\%}              & {\color[HTML]{000000} 10.87\%}            & {\color[HTML]{000000} 2.93\%}             & {\color[HTML]{000000} 36.05\%}           \\ \hline

\end{tabular}
}
\end{footnotesize}
\end{table}

Deepmerge is known to be sensitive to the input merge conflict size as it performs better in small merges.
This is possible because the input to the model is carefully tuned and it is represented token by token. 
Instead, in \app, the input is aligned in a sequence of variant A, variant B and base O and fed into the language model all at once.
This way of combining the inputs of A, B and O is termed as \lstinline{naive} in DeepMerge.
When using the same input representation, \app is slightly outperformed (15.02\% vs 14.09\%) the Deepmerge model.

In addition, using the classic conflict formulation in the prompt improve the result by 25\% (11.99\% vs 15.02\%).
This result shows that having the right prompt format improves the resolution accuracy. 

\subsection{Discussion}
In summary, these two case studies show that \app has modest to competitive performance on problems where current SOTA approaches require special-purpose domain specific languages (DSL) for program synthesis or fine-tuning a neural network model (that requires tens of thousands of data points).
It highlights that language models still do not obviate the need for domain-specific investments for the merge conflict problem. 
Further, even when the data curation is not feasible due to the lack of repository information, prompt engineering is still useful to improve the accuracy of the resolution in \app.
\section{Related Work}
\label{sec:related}

\para{Semantics Merge Conflict}
Semantics merge conflicts happen when the merged code cannot be successfully compiled.
This problem was first introduced by Horwitz {\it et .al}~\cite{Horwitz88} and later formalized by Yang {\it et .al}~\cite{Wuu90} in 1990s.
A recent study~\cite{Souze03,Ghiotto18} has shown that having such a bad merge in the code delayed the development cycle or caused damage by simply leaving bugs in the code.
As a result, semantics merge conflict detection~\cite{Sousa18} and resolution approaches~\cite{Shao09,SungLKCW20} have been proposed.
JDIME~\cite{JDIME} automatically tunes a semi-structured merge based conflict locations detection and resolution.
NLX\_REG~\cite{rahmani2021} uses large language model to synthesizing regular expressions.
SAFEMERGE~\cite{Sousa18} prevent merge conflicts by defining formal specifications to the based code, variants of the code and the final merged code.
However, it did not directly produce the merge conflict resolutions as \app does.

The closest work to ours is the MrgbldBrkFixer~\cite{SungLKCW20}.
It resolves semantics merge conflict for a divergent fork by analyzing the AST diffs for changes in the upstream to construct a patch for merge conflicts. 
However, MrgbldBrkFixer still requires developers' manual work to classify the build breaks, and the tool heavily relies on the AST analysis for C++ code only.
In contrast, \app is scalable, fully automated and language-agnostic by leveraging large scale language models.

\para{Textual Merge Conflict} 
Textual merge conflicts have been long known as a severe and challenging problem, as reported in prior studies~\cite{Ghiotto18,PanLNGLK21}. 
As a result, textual merge conflict mining and detection approaches~\cite{Brun11,Guimaraes12,Nguyen2015DetectingSM,JDIME} have been proposed.
Going one step further than bad merge prevention, to directly resolve the merge conflict, recently, we have witnessed great progress via program synthesis\cite{PanLNGLK21} and machine learning~\cite{dinella2021deepmerge}.
Deepmerge~\cite{dinella2021deepmerge} required customized machine learning models.
Pan {\it et .al}\cite{PanLNGLK21} studied the historical data of bad merge to design 
special purpose domain-specific languages (DSL) for program synthesis.
IntelliMerge~\cite{IntelliMerge19} studied refactoring caused merge conflicts in software development and evolution in Java programs.


\section{Conclusion}

In this paper, we explored the feasibility of leveraging k-shot learning with large language models for resolving various merge conflicts (both textual and semantic).
Our results demonstrate that LMs have the potential to be useful for this important problem in software engineering by providing cost-effective solutions ranging from SOTA performance on some domains (e.g. in semantic merge conflicts in Edge), while providing  modest performance (on textual conflicts) on other domains.
Our work also illustrates the importance of prompt engineering for these language models as an important avenue for research, including automating the most effective prompts given data from a domain.

In future work, we are working on feeding more structured and merge-aware representation of inputs in the prompts to better communicate the problem to the LMs.
For instance, one can perform an edit-aware encoding of the merge input as done in DeepMerge, or encode semantics of pointer networks to only rearrange lines from a conflicted region. 


\bibliographystyle{ACM-Reference-Format}
\bibliography{msr_intern}

\end{document}